\documentclass[journal=jacsat,manuscript=article]{achemso}
\SectionNumbersOn
\usepackage{amsmath,amscd,amsfonts,amssymb,color}
\usepackage{graphicx}
\usepackage{epstopdf}
\usepackage{pifont}
\usepackage{xcolor}
\usepackage{hyperref}
\newcommand{\eg}{\textit{e.g. }}
\newcommand{\ie}{\textit{i.e. }}

\author{Bindu\footnote{These authors contributed equally to this work.}}
\affiliation{Institute of Applied Physics, The Hebrew University of Jerusalem, Jerusalem, 9190401, Israel}

\author{Amandeep Singh$^a$}
\email{www.amansidhu.com@gmail.com}
\affiliation{Institute of Applied Physics, The Hebrew University of Jerusalem, Jerusalem, 9190401, Israel}

\author{Amir Hen}
\affiliation{Institute of Applied Physics, The Hebrew University of Jerusalem, Jerusalem, 9190401, Israel}

\author{Lukas Drago \'{C}avar}
\affiliation{Institute of Physics, Johannes Gutenberg University Mainz, 55128 Mainz, Germany}

\author{Sebastian Maria Ulrich Schultheis}
\affiliation{Institute of Physics, Johannes Gutenberg University Mainz, 55128 Mainz, Germany}

\author{Shira Yochelis}
\affiliation{Institute of Applied Physics, The Hebrew University of Jerusalem, Jerusalem, 9190401, Israel}

\author{Yossi Paltiel}
\affiliation{Institute of Applied Physics, The Hebrew University of Jerusalem, Jerusalem, 9190401, Israel}

\author{Andrew F. May}
\affiliation{Materials Science and Technology Division, Oak Ridge National Laboratory, Oak Ridge, TN 37831, USA}

\author{Angela Wittmann}
\affiliation{Institute of Physics, Johannes Gutenberg University Mainz, 55128 Mainz, Germany}

\author{Mathias Kl\"{a}ui}
\affiliation{Institute of Physics, Johannes Gutenberg University Mainz, 55128 Mainz, Germany}

\author{Dmitry Budker}
\affiliation{Johannes Gutenberg-Universität Mainz, 55128 Mainz, Germany}
\altaffiliation{Helmholtz-Institut Mainz, GSI Helmholtzzentrum für Schwerionenforschung, 55128 Mainz, Germany}
\altaffiliation{Department of Physics, University of California, Berkeley, CA 94720, USA}

\author{Hadar Steinberg}
\affiliation{The Racah Institute of Physics, The Hebrew University of Jerusalem, Jerusalem 9190401, Israel}
\altaffiliation{The Center of Nano-Science and Nanotechnology, The Hebrew University of Jerusalem, Jerusalem, 9190401, Israel}

\author{Nir Bar-Gill}
\email{bargill@phys.huji.ac.il}
\affiliation{Institute of Applied Physics, The Hebrew University of Jerusalem, Jerusalem, 9190401, Israel}
\alsoaffiliation{The Racah Institute of Physics, The Hebrew University of Jerusalem, Jerusalem 9190401, Israel}

\title{Quantum Imaging of Ferromagnetic van der Waals Magnetic Domain Structures at Ambient Conditions}

\begin{document}
\pagebreak
\begin{abstract}
Recently discovered 2D van der Waals magnetic materials, and specifically Iron-Germanium-Telluride ($\rm Fe_{5}GeTe_{2}$), have attracted significant attention both from a fundamental perspective and for potential applications. Key open questions concern their domain structure and magnetic phase transition temperature as a function of sample thickness and external field, as well as implications for integration into devices such as magnetic memories and logic. Here we address key questions using a nitrogen-vacancy center based quantum magnetic microscope, enabling direct imaging of the magnetization of $\rm Fe_{5}GeTe_{2}$ at sub-micron spatial resolution as a function of temperature, magnetic field, and thickness. We employ spatially resolved measures, including magnetization variance and cross-correlation, and find a significant spread in transition temperature yet with no clear dependence on thickness down to 15 nm. We also identify previously unknown stripe features in the optical as well as magnetic images, which we attribute to modulations of the constituting elements during crystal synthesis and subsequent oxidation. Our results suggest that the magnetic anisotropy in this material does not play a crucial role in their magnetic properties, leading to a magnetic phase transition of $\rm Fe_{5}GeTe_{2}$ which is largely thickness-independent down to 15 nm. Our findings could be significant in designing future spintronic devices, magnetic memories and logic with 2D van der Waals magnetic materials.
\end{abstract}

\section{Introduction}
In recent years, the advent of 2D van der Waals (vdW) materials that can be exfoliated down to one or few layers has opened new frontiers encompassing material science, fundamental physics, and novel applications \cite{liu-nrm-16,frisenda-npj-20}. Magnetic vdW materials form an important subclass of this family \cite{Li-AM-19}, posing significant open questions, both fundamental and applied. For example, unknown aspects of these materials are associated with the structure of magnetization in thin flakes, thickness dependence of their magnetic properties, and implications of interfacial and structural properties on magnetization.

Studies of magnetic vdW materials have addressed some of these questions, spanning different materials and different experimental techniques. These include anomalous Hall effect measurements \cite{Deng-NL-22}, reflective magnetic circular dichroism \cite{Meng-NC-21}, magneto-optical Kerr effect \cite{Huang-Nature-17}, and integration of magnetic barriers into tunneling devices \cite{Klein-science-18}, addressing the behavior of magnetization, magnetic domains, and Curie temperature ($\rm T_C$) in recently discovered 2D vdW magnets. Moreover, magnetic imaging techniques have been employed to obtain spatially resolved information related to the magnetic properties of such materials\cite{chen-ami-23,thiel-sci-19,Yi-2Dmaterial-16}. Specifically, nitrogen-vacancy (NV) center based magnetic imaging modalities, both wide field imaging \cite{McLaughlin-NL-22,Broadway-AM-20} and scanning-tip microscopy \cite{Tschudin-NC-24,thiel-sci-19} have been employed in this context. 

In this work, we focus on the recently identified vdW ferromagnet $\rm Fe_5GeTe_2$ \cite{stahl-zfauac-18, may-acs-19}, due to its near-room-temperature $\rm T_C$ \cite{chen-IOP-22} and its potential relevance for future research and applications. Bulk and thin layer studies of this material have characterized its magnetization structure and $\rm T_C$ \cite{may-acs-19}, current-induced domain-wall motion \cite{Robertson-IOP-23}, and ferromagnetic resonance \cite{Alahmed-IOP-21}.

Here we go beyond previous results and use NV center magnetic microscopy to investigate the impact of interfacial and anisotropy effects on vdW magnets. We study the magnetization structure of FGT for various flake thicknesses, as a function of external magnetic field and temperature. Through spatial magnetic imaging of selected flakes, we identify a large spread in $\rm T_C$ regardless of flake thickness, down to 15 nm. We further find previosuly unknown magnetic structures associated with crystallographic features in the material. These results provide important information on the magnetic properties of FGT and vdW magnets in general, highlighting the relevance of interfacial effects on magnetic behavior.
%
%
\subsection{Quantum Magnetic Sensing}
Quantum sensing in general is a highly developed field, based on the premise of using a quantum system to sense physical quantities such as magnetic fields, temperature and more \cite{degen-rmp-17, Budker-NP-07}.  A broad range of quantum sensor implementations have been realized, including neutral atoms\cite{kitching-IEEE-11}, trapped ions\cite{Maiwald-NP-09}, solid-state spins \cite{Bar-Gill-NC-13}, photons\cite{Pirandola-NP-18}, and superconducting circuits \cite{drung-IEEE-07}.
    
A significant modality of quantum sensing addresses magnetic field sensing, with applications ranging from biomedical to material science. Various techniques have been developed and employed for magnetic field sensing and imaging, \eg superconducting quantum interference device (SQUID) \cite{ya-RMP-98, drung-IEEE-07}, magnetic resonance force microscopy \cite{rugar-nature-04, degen-NAC-09}, scanning Hall probe microscopy \cite{chang-APL-92, shaw-RSI-16}, and optical atomic magnetometers \cite{ledbetter-NAS-08, Xu-NAS-06} to name a few. These techniques differ in their advantages and disadvantages, offering some combination of high sensitivity and spatial resolution, but sometimes require high vacuum and/or cryogenic temperature to operate. 

Quantum sensors based on crystal solid-state defects are promising candidates to circumvent such limitations. NV centers in diamond \cite{dyer-JEAP-65}, boron-vacancies in hexagonal boron nitride \cite{tran-natnan-16}, silicon-vacancy in diamond and silicon-carbide vacancies \cite{nagy-natcom-19} are leading examples of such solid-state defects. Amongst various solid-state defects, NV centers have been widely explored for room temperature magnetic sensing \cite{barry-prl-20}.  NV centers with long spin coherence time \cite{Bar-Gill-NC-13} can be optically initialized, manipulated, and read out under ambient conditions \cite{hanson-PRB-06}. Further, the optical transitions of NV centers are highly sensitive to physical conditions such as temperature \cite{sunuk-CAP-18}, pressure \cite{hilberer-prb-23}, electric field \cite{Dolde-NP-11} and magnetic field \cite{Bar-Gill-NC-13} which makes them well suited for quantum sensing.
    
NV center based magnetometers demonstrated femto-$\rm T{Hz}^{-1/2}$ order magnetic sensitivity \cite{xie-SB-21}. Such magnetometers are typically used for magnetic field sensing and imaging in fundamentally two different configurations. In a wide-field setup,  a quasi-2D dense NV ensemble closer to the diamond surface serves as a 2D magnetic sensor. One may directly transfer the magnetic sample onto the diamond's sensing surface which encodes the sample's magnetic features to the localized NV centers in the proximity of the magnetic sample. From the optical readout, one can extract the magnetic image having diffraction-limited spatial resolution \cite{scholten-JAP-21}. This configuration is sometimes referred to as a Quantum Diamond Microscope (QDM). Alternatively, a scanning single NV center technique can be employed to further enhance the spatial resolution down to a few nanometers \cite{degen-APL-08}. Both configurations are in use based on the application. NV center based techniques find applications in biomedical science \cite{zhang-ACS-21, aslm-nrp-23} for magnetic imaging of living cells \cite{LeSage-Nature-13}, magnetic field sensing in artificial magnetic nanoparticles for drug delivery \cite{manuel-NT_07, omid-ADDR-10}, monitoring the drug efficacy \cite{Arturo-NL-23} and sensing free radical generation in cells \cite{Fan-ACS-24}. NV centers were also employed for sensing in condensed matter physics \cite{Casola-NRM-18}, \eg for understanding the charge, spin, and magnetic behavior at the nanoscale in the recently discovered 2D vdW materials \cite{Geim-Nature-13} and topological insulators \cite{hasan-RMP-10}. 
%
%
\subsection{Magnetic vdW Materials}
The discovery of 2D vdW materials has opened a broad range of novel research fields \cite{novoselov-sci-16}, as well as new avenues for applications, such as next-generation computational and spintronic devices \cite{Ahn-npj-20}, and has been extensively explored in the past decade. A variety of 2D materials including ferromagnets \cite{thiel-sci-19}, antiferromagnets \cite{li-nl-23}, superconductors \cite{Xi-NP-16}, insulators \cite{Mclaughlin-AQT-21}, and semiconductors \cite{Manzeli-NRM-17} have been studied from cryogenic to room temperature, to understand their magnetic response at the nanoscale with the aim of identifying the different behavior of the 2D material from its original  bulk crystal \cite{Casola-NRM-18}.

One of the vdW magnets of interest is iron-germanium-telluride $\rm Fe_{(5-x)}GeTe_2$ (FGT), due to its near room temperature T$\rm _C$ \cite{may-acs-19}, making it a suitable candidate for real-world nano-spintronic applications. FGT is a cleavable 2D vdW ferromagnet with long-range ferromagnetism originating from significant spin polarization of delocalized ligand Te states \cite{Yamagami-PRB-21}. FGT exhibits a tunable magnetic phase transition between ferromagnetism and antiferromagnetism by applying gate voltage, making it a suitable candidate for fast and energy-efficient data storage devices \cite{Tan-NL-21}. In addition to using 2D magnetic materials for information transfer, spin structures such as magnetic skyrmions, domain-walls, and spin waves can serve as information carriers. In this context, FGT hosts these spin structures at room temperature \cite{Schmitt-CP-2022}. 

Due to the reduced dimensionality, these 2D magnetic materials have unique properties depending upon their growth conditions and thicknesses compared to their bulk counterparts. In thin layers, the geometric shape factor promotes in-plane magnetic anisotropy and this competes with out-of-plane magnetic anisotropy (or enhances magnetism that intrinsically has a preference for in-plane moment orientation). In some materials, the competing anisotropies lead to the stabilization of magnetic phases in thin flakes (or films) that are not present in bulk crystals \cite{Fei-NM-18,Huang-PRB-94}.

Here, we image and study FGT magnetization under various conditions, including temperature and external magnetic field, with a focus on the effect of the thickness of FGT flakes on their T$\rm _C$ and magnetization structures. High-resolution NV center based magnetic imaging reveals domain structures and correlations, specifically relating magnetization with large-scale crystallographic effects. These results, along with the variations identified in the T$\rm_C$ of flakes with thicknesses ranging between 15-220 nm, provide fundamental insights into the magnetization in these materials, and anisotropy and interfacial effects, with implications on integrated spintronic devices for specific applications, such as magnetic memories and logic.
%
%
\section{Experimental Setup}\label{ExpSetup} 
\begin{figure}[H]
\begin{center}
\includegraphics[angle=0,scale=0.95]{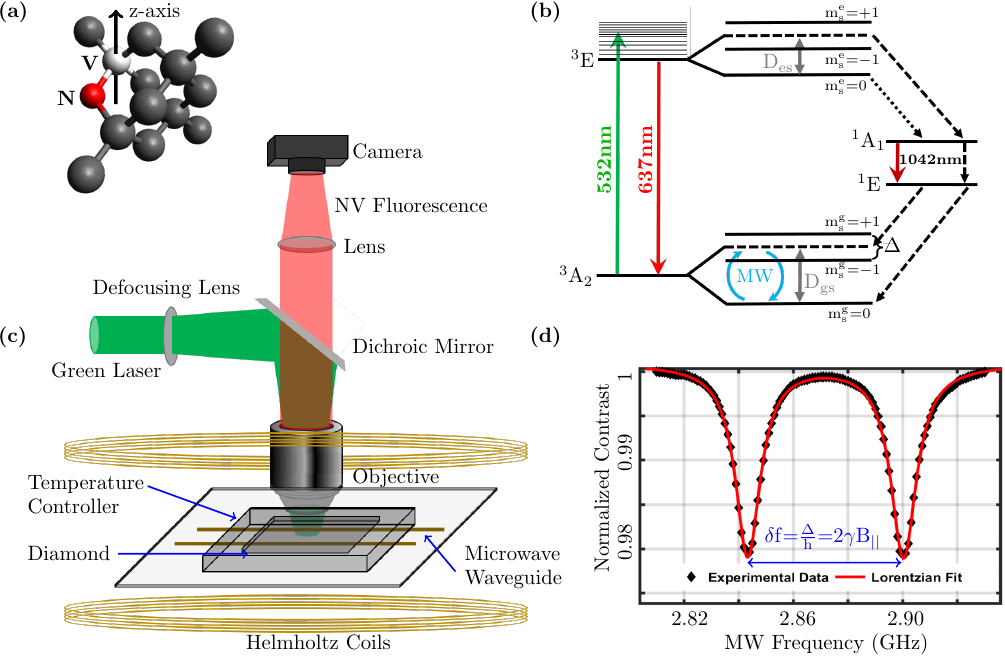}
\end{center}
\caption{(a) Diamond lattice hosting a nitrogen (red) vacancy (white)  defect. Black balls represent carbon atoms. An external bias magnetic field is applied along the z-axis and has equal magnitudes of the projections on each of the four possible NV center axes. (b) The spin-1 NV center energy level diagram depicting optical pumping with a 532 nm green laser from the ground triplet ($^3$A$_2$) to the excited triplet ($^3$E) which may decay via various channels. Dotted/dashed arrows represent optically forbidden transitions. Each spin state is labeled with the respective magnetic spin quantum number, m$_s$, $\Delta$ represents the Zeeman splitting. $\rm D_{gs}$ = 2.87 GHz is the ground state zero-field splitting, excited state zero-field splitting $\rm D_{es}$ = 1.42 GHz and MW represents the microwave control. (c) Schematic of the experimental setup of NV center quantum magnetometer. (d) A typical ODMR under an external bias magnetic field. Black diamonds ($\blacklozenge$) are the experimental points while the red line (\textcolor{red}{\huge \textbf{-}}) is a Lorentzian fit.}
\label{setup}
\end{figure}

The NV sensor used is based on a 3 mm$\times$3 mm$\times$0.1 mm single crystal electronic grade diamond from Element Six. The diamond was implanted with a 2$\times$10$\rm ^{13}$ $\rm cm^{-2}$ dose of 10 keV nitrogen (N$\rm ^{15}$) ions, followed by annealing and electron irradiation to create an NV center ensemble layer $\sim$20 nm below the diamond surface. The diamond was cut perpendicular to the crystallographic x-axis \ie along the $\rm \{1 0 0\}$ plane \cite{kittel-book-04}. The schematic of a diamond lattice hosting an NV center defect is depicted in Fig.\ref{setup}(a). In this lattice configuration, a bias magnetic field along the z-axis has equal projections on all the four possible NV center axes, to be specific $ \rm B_{||}=B_z^{ext} \cdot cos(54.75^{\circ})$, where $\rm B_{||}$ is the component of B$\rm ^{ext}_z$ parallel to NV center axis.
%
%
%
%

Fig.\ref{setup}(b) depicts a typical energy level diagram of an NV center. A green laser (532 nm) initializes the NV center ensemble in the $ \vert \rm m_s^g=0 \rangle$ or simply $\vert 0 \rangle$ state. This is achieved by optically pumping the population of the NV center ensemble from the ground triplet ($^3$A$_2$) state to the excited triplet ($^3$E) states. While the system is in the excited triplet ($^3$E) state, it has several decay paths with different probabilities. Optical transitions can take the NV back to the ground triplet ($^3$A$_2$) and are spin preserving. Non radiative transitions to the intermediate singlet state ($^1$A$_1$) are shown by dashed/dotted arrows in Fig.\ref{setup}(b). The $m_s^e=\pm 1$ states have a higher decay probability (dashed arrow, Fig.\ref{setup}(b)) than the $m_s^e=0$ state (dotted arrow, Fig.\ref{setup}(b)) to the singlet state $^1$A$_1$. Hence the spin preserving optical relaxation from the excited triplet ($^3$E) to the ground triplet ($^3$A$_2$) is suppressed for $m_s=\pm 1$, which in turn builds up the $\vert 0 \rangle$ population. Both radiative and non-radiative decays, from the $\rm ^1A_1$ to the $\rm ^1E$ singlet, are possible while non-radiative decay from $\rm ^1E$ singlet to the $\rm ^3A_2$ triplet has identical probabilities for $\vert 0 \rangle$ and $\vert \pm 1 \rangle$ (marked by the dashed arrow in Fig.\ref{setup}(b)). An optical transition from the excited triplet state to the ground triplet state gives rise to a characteristic fluorescence at a wavelength of 637 nm \ie the zero phonon line, with a broad phonon sideband (637-800 nm). 

The experimental setup Fig.\ref{setup}(c) includes (in addition to the optics) microwave (MW) delivery, which can induce transitions between the levels $\vert 0 \rangle$ and $\vert \pm1\rangle$. For the population distribution of the optically pumped equilibrium state, any coherent manipulation between the $\vert 0 \rangle$ and $\vert \pm1\rangle$ levels using MW drive, results in a decreased fluorescence as depicted in Fig.\ref{setup}(d). In the absence of any external bias magnetic field, one would expect the lowest fluorescence, hence extremum normalized contrast, when the MW drive frequency is swept around $\rm D_{gs}$=2.87 GHz, \ie the zero-field splitting (ZFS) between $\vert 0 \rangle$ and $\vert \pm1\rangle$.

An external bias field Fig.\ref{setup}(c) in the z-direction, typically 30-35 Gauss, is applied using Helmholtz coils to remove the degeneracy of $\vert \pm1\rangle$ which results in Zeeman splitting of the NV center ground state manifold by an amount $\rm \Delta=2\gamma B_{||}$, Fig.\ref{setup}(b), where $\gamma$ is the electronic gyro-magnetic ratio. We note that B$_{\perp}$, \ie the component of the external bias magnetic field perpendicular to the NV center axis, competes with the ZFS, and its effect can be neglected on the Zeeman splitting under first-order approximations \cite{thiel-sci-19}.  A coplanar MW waveguide is used to coherently control the transitions between the levels $\vert 0 \rangle$ and $\vert\pm1\rangle$. Experiments were performed at room temperature. To further control the temperature of the sample under investigation, in the range $5^{\circ}$- $55^{\circ}$ C (\ie 278-328 K), a Peltier temperature controller was employed.

Finally, air objectives with 40 and 60 magnifications, having numerical apertures (NA) 0.65 and 0.95 respectively, were used for excitation and collection of the NV center fluorescence. The collected fluorescence was recorded using an Andor Neo 5.5 sCMOS camera, with 2560$\times$2160 active pixels (5.5 Megapixels). Depending upon the objective magnification the pixel size, with an appropriate camera binning, is generally of the order of the diffraction limit $\rm \approx\lambda/2NA \sim300$ nm. It is worth noting that the experiment is designed such that the camera records the fluorescence pixel-wise, for the duration of the exposure time, at each MW frequency. One can extract the pixel-wise optically detected magnetic resonance (ODMR) for the chosen region of interest (ROI). To compute the contrast at a certain MW frequency one looks for the change in the fluorescence with and without MW drive. This is followed by averaging the ODMR data over all the pixels in the ROI to generate the resulting ODMR plot. A typical ODMR plot under a bias field is shown in Fig.\ref{setup}(d).
%
%
\section{Quantum Magnetic Sensing with NV Centers}\label{MagSen}
An external bias magnetic field removes the degeneracy of the $\vert \pm1 \rangle$ manifold. The Zeeman splitting between $\vert +1 \rangle$ and $\vert - 1 \rangle$ \ie $\rm \Delta=2\gamma B_{||}$, can readily be measure experimentally from the ODMR, Fig.\ref{setup}(d), which eventually can be used to compute B$_{||}$ and hence B$\rm _z$. In an actual sensing experiment, B$\rm _z$ could be composed of the applied bias field B$\rm ^{ext}_z$ and the stray field to be sensed. The stray z-field detected by the QDM will be denoted by B$\rm _z$. As the experimental setup is designed to acquire the pixel-wise ODMR, one may experimentally find the magnetic field seen by each pixel with a spatial resolution limited by the diffraction limit.
%
%
\subsection{Magnetic Sensitivity}
To evaluate the magnetic sensitivity of a magnetic sensor, similar to what is employed here, one may typically be interested in computing the signal-to-noise (SNR) limited minimum detectable magnetic field ($ \rm \delta B_{min}$) from an ODMR spectrum \cite{dreau-prb-11}.  It can be obtained using the expression:
\begin{equation}\label{sensitivity}
\rm \delta B_{\min} = \frac{\delta\beta}{\gamma \cdot \left(\frac{\partial\beta}{\partial\nu}\right)_{max}}
\end{equation}
where $\rm \delta \beta$ is the signal noise at the point of maximum slope $\left(\frac{\partial\beta}{\partial\nu}\right)_{max}$ of the ODMR spectrum. The sensitivity, $\eta$, for the detection duration $t$ is given by $\rm \eta = \delta B_{\min} \cdot \sqrt{t}$ \cite{dreau-prb-11}. The sensitivity, corresponding to the ODMR shown in Fig.\ref{setup}(d), is 172.3 $\pm$6.8 $ \rm nT/\sqrt{Hz}$.
%
%
\subsection{FGT Sample}\label{sample}
Identifying and synthesizing a 2D vdW magnetic material with room temperature T$\rm_C$  is challenging. After the discovery of ferromagnetism in Fe$_3$GeTe$_2$ with T$_C$ $\sim$220 K, efforts were made to identify related van der Waals ferromagnets with higher ordering temperatures by employing thicker Fe-Ge slabs into the structure.  Such efforts resulted in the discovery of Fe$_{(5-x)}$GeTe$_2$ materials with ferromagnetism reported near room temperature \cite{stahl-zfauac-18}. Recent work showed that with suitable doping using cobalt \cite{may-prm-20,tian-apl-20} or nickel \cite{chen-PRL-22} one can reach above room temperature T$\rm_C$.

The sample used in the current study is a cleavable ferromagnetic 2D vdW crystal Fe$_{(5-x)}$GeTe$_2$. The crystals were quenched as described in \cite{stahl-zfauac-18} and directly used without further processing; the experimental value of x is approximately 0.3 based on wavelength dispersive spectroscopy as reported in \cite{stahl-zfauac-18}. Supporting Information Section-S1 \cite{si-ACS-25} details the process for exfoliation and transfer of FGT flakes. Depending upon the iron content `x', the FGT bulk single crystal has a T$\rm_C$  ranging from 260 K to 310 K \cite{Alahmed-IOP-21} and retains magnetism down to a few nm thickness\cite{may-acs-19}. In contrast, depending upon the flake thickness, the T$\rm_C$  for 2D flakes ranges between 270 K and 300 K \cite{may-acs-19,chen-ami-23}. It is interesting to note that the T$\rm_C$  not only depends upon the thickness but also on other factors such as the thermal and magnetic history of the bulk material, chemical and mechanical properties of the crystal, etc. Typically, the FGT flakes are observed to have an out-of-plane (OOP) magnetization. The induced magnetization pattern depends upon the bias field strength and temperature \cite{may-acs-19,chen-ami-23,antonio-jap-21}. This OOP magnetization makes FGT a good candidate to study under only a z-oriented bias field (even though our QDM setup is designed to extract arbitrarily directed magnetization).

%
%
\begin{figure}
\includegraphics[scale=1]{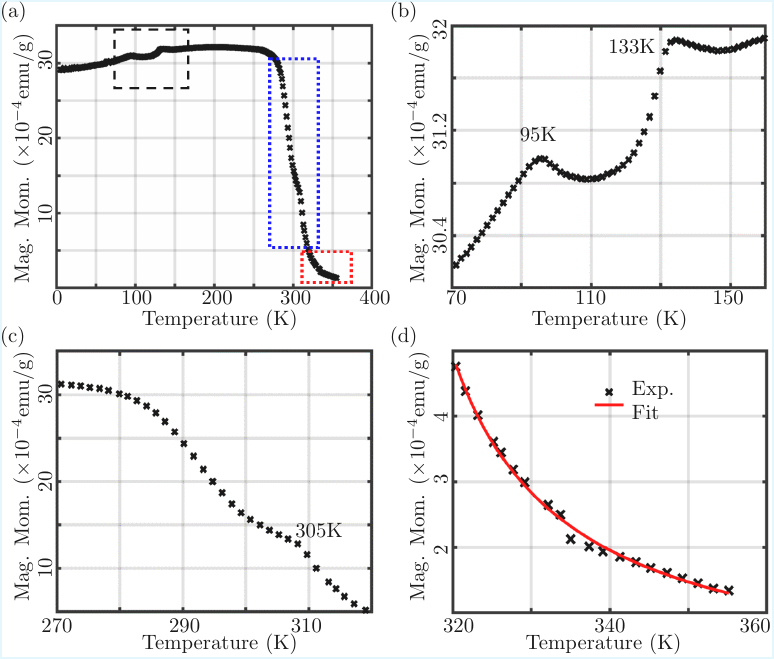}
\caption{Magnetization characterization of bulk FGT single crystal. (a) Zero field cooled magnetic moment as a function of the temperature under 100 Oersted magnetic field. (b)  A zoomed-in plot of the black dashed square in (a) showing two transitions at 95 K and 133 K. (c) A zoomed-in plot of the blue dashed rectangle in (a) showing ferromagnetic to paramagnetic phase transition at $\sim$305K. (d) A zoomed-in plot of the dotted red rectangle in (a) showing the paramagnetic phase fitted using Curie-Weiss law giving T$\rm_C$=305.5$\pm$2.5 K.}
\label{Bulk Tc plot}
\end{figure}
%
%
\subsection{Bulk FGT Measurements}\label{Bulk Measurements}
To determine the T$\rm_C$ of the parent bulk single crystal FGT used in this work, the magnetic moment was measured using a Quantum-Design SQUID magnetometer as a function of temperature under an external magnetic field of 100 Oersted, with zero field cooling (ZFC). Zero field cooling implies that the FGT bulk sample was cooled to near zero Kelvin under zero magnetic field. Once cooled, a 100 Oersted magnetic field was applied to magnetize the sample and the magnetization was recorded with increasing temperature. The results are depicted in Fig.\ref{Bulk Tc plot}, revealing a pair of transitions at around 95 K and 133 K which are visible in the zoomed-in plot of Fig.\ref{Bulk Tc plot}(b).  

Upon warming in the ZFC condition, the magnetization of the bulk FGT reaches a maximum between 170 K and 250 K. At higher temperatures ($\sim$260 K), the magnetization starts decreasing, and as the temperature further increases, the FGT transitions from the ferromagnetic to paramagnetic phase, Fig.\ref{Bulk Tc plot}(c). One may observe a kink at $\sim$305 K displaying the magnetic phase transition of bulk FGT.  Above T$\rm_C$, the ZFC branch exhibits the typical paramagnetic shape and fitting the data by the Curie-Weiss law (red curve in Fig.\ref{Bulk Tc plot}(d)) given by the equation \cite{Danielian-pps-62,rhodes-prsl-63}
\begin{equation}
   \rm \chi=\frac{C}{T-\Theta}
\end{equation}
$\chi$ being magnetic susceptibility, C is the material specific Curie's constant and $\Theta$ is the Weiss constant \cite{rhodes-prsl-63} typically larger than Curie temperature. This yields $\Theta$ = 305.5$\pm$2.5 K, a value which fits well with T$\rm_C$ mentioned above, Fig.\ref{Bulk Tc plot}(c).
%
%
\section{Quantum Magnetic Imaging of FGT Flakes}\label{Flakeimage1}
\begin{figure}
\includegraphics[angle=0,scale=0.95]{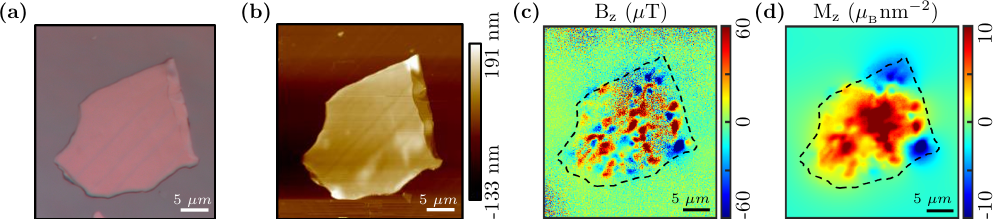}
\caption{(a) Optical image of an FGT flake transferred onto the diamond surface by mechanical exfoliation, see Supporting Information Section-S1 \cite{si-ACS-25} for detail. (b)  AFM topography of the FGT flake. (c) The extracted stray magnetic field, sensed by QDM, of the FGT flake under a 1800 $\mu$T z-bias field at 288 K. (d) The reconstructed OOP magnetization ($\rm M_z$) \cite{thiel-sci-19, tan-IEEE-96} from the OOP stray magnetic field ($\rm B_z$) of the flake shown in (c).}
\label{flake0}
\end{figure}

Under ambient conditions, we study FGT flakes utilizing our in-house developed wide-field NV center based QDM. An optical image at 100X magnification of one such mechanically exfoliated flake, see Supporting Information Section-S1 \cite{si-ACS-25} for detail, transferred onto the diamond surface is shown in Fig. \ref{flake0}(a). The flake's average thickness was measured as 137 nm using an atomic force microscope (AFM) (Supporting Information Section-S2 \cite{si-ACS-25}). The AFM topograph of the FGT flake is shown in Fig. \ref{flake0}(b). The extracted stray magnetic field image from the pixel-wise ODMR data is shown in Fig. \ref{flake0}(c). The experiment was performed at 288 K and $\sim$1800 $\mu$T z-bias magnetic field. The QDM reveals sub-micrometer scale magnetic domain features indicating the presence of ferromagnetism and a phase transition expected at room temperature or above for this material. The typical average domain width for the bulk FGT is reported $\approx$250 nm at 50 K utilizing X-ray photo-emission electron microscopy (XPEEM). In addition, the usual domain structure with well-defined domain-walls ceases to exist as the thickness of the FGT flakes decreases. XPEEM-based analysis to visualize the OOP magnetic field for several layers of FGT was reported recently \cite{fujita-acs-22}.
%
%
%
%

Further, to reconstruct the OOP magnetization ($\rm M_z$), corresponding to the measured OOP B$\rm _z$ of the FGT flake shown in Fig.\ref{flake0}(c), the method of reverse propagation of magnetization was utilized \cite{tan-IEEE-96, thiel-sci-19}. The reverse propagation approach allows for the reconstruction of a 2D magnetization map from the measured 2D stray magnetic field. The details of the method used for computing M$\rm _z$ given the B$\rm _z$ map of a 2D source, based on the proposals in Ref. \cite{tan-IEEE-96}, can be found in the supplementary material of \cite{thiel-sci-19}. Fig. \ref{flake0}(d) depicts the reconstructed 2D magnetization image of the FGT flake. Supporting Information Section-S3 \cite{si-ACS-25} describes the steps required for this calculation \cite{tan-IEEE-96,thiel-sci-19}.
%
%
\subsection{Temperature Dependence of Stray Fields}\label{temp depend}
For insights into the phase transition and to compute the transition temperature (T$\rm_C$ ), we performed experiments characterizing the temperature dependence of the B$\rm _z$ of the FGT flakes. A feedback-based Peltier temperature controller, with $\pm$0.1 K precision, was employed to control the temperature of the flakes (Supporting Information Section-S4 \cite{si-ACS-25}). In a typical temperature-dependent measurement of B$\rm _z$, the temperature was varied between 278 K and 328 K.

\begin{figure}[H]
\includegraphics[scale=0.85]{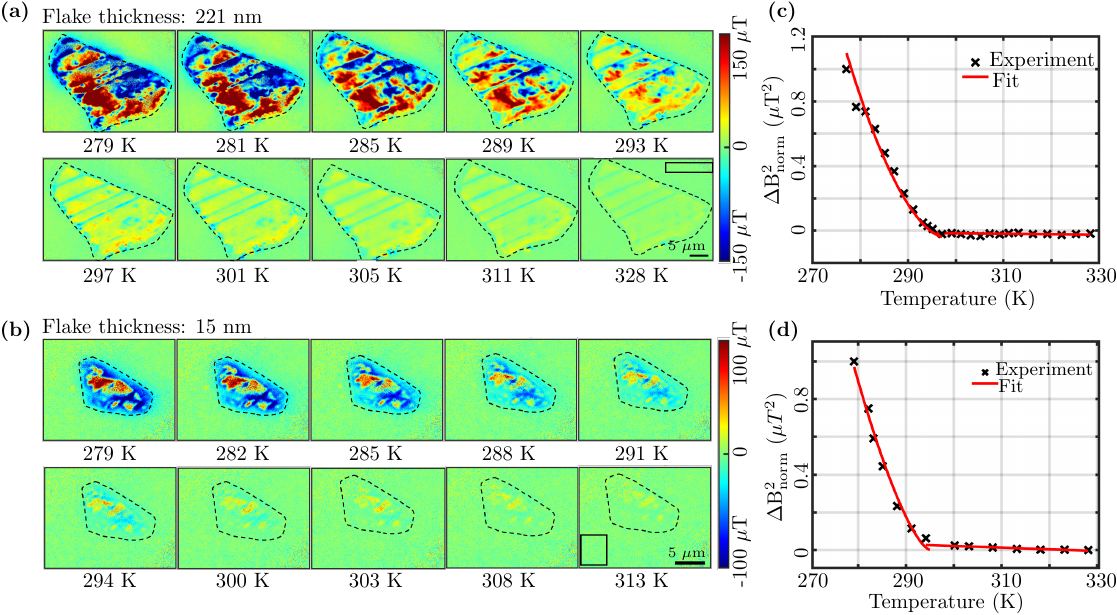}
\caption{Temperature-dependent quantum magnetic images and phase transition plots. (a) Experimentally extracted OOP stray magnetic field (B$\rm _z$) of 221 nm thick FGT flake under 3150 $\mu$T z-bias field for the temperature in range 279-328 K. In the magnetic image, one can observe the multiple domains revealing ferromagnetic ordering at low temperatures. As the temperature increases, domains start rearranging, eventually disappearing at around 300 K, displaying the magnetic phase transition. (b) Experimentally extracted OOP stray magnetic field (B$\rm _z$) of a 15 nm thin FGT flake under 3150 $\mu$T z-bias field for the temperature in range 279-313 K. (c) The stray magnetic field variance $\Delta$B$\rm ^2_{norm}$ versus temperature plot for the FGT flake shown in (a). The retrieved T$\rm_C$ is 296$\pm$1.5 K. (d) The stray magnetic field variance $\Delta$B$\rm ^2_{norm}$ versus temperature plot for the FGT flake shown in (b). The retrieved T$\rm_C$ is 295$\pm$2.1 K. Black rectangles in (a) and (b) are the background and the dashed line encapsulating the flakes are the regions considered for computing the normalized stray field variance $\Delta$B$\rm ^2_{norm}$.}
\label{flake_1}
\end{figure}

Figure \ref{flake_1} depicts one such temperature-dependent measurement of the stray magnetic fields for two different FGT flakes with 221 nm (thick) and 15 nm (thin) thicknesses. Several representative images of the FGT flake, at different temperatures, are shown here (additional measured magnetic maps are given in  Supporting Information Section-S6 \cite{si-ACS-25}). In Figs. \ref{flake_1}(a) and (b), the temperature-dependent behavior of the B$\rm _z$ can be observed. One may notice the sub-micrometer-sized magnetic domains. At lower temperatures, domains have stray fields of the order of 250 $\mu$T at a distance of $\sim$20 nm. With the increase in temperature, domains seem to rearrange and appear to shrink during the phase transition and then completely vanish at elevated temperatures. The phase transition of 2D FGT flakes from ferromagnetic to paramagnetic phase is characterized by a near room-temperature T$\rm_C$ \cite{may-acs-19, chen-ami-23}.

For the phase transition plot of the FGT flakes, to extract T$\rm_C$, the normalized stray field variance ($\Delta$B$\rm ^2_{norm}$) \cite{chen-ami-23} is calculated. The $\Delta$B$\rm ^2_{norm}$ versus temperature plot for the thick and thin flakes are shown in Fig. \ref{flake_1}(c) and (d) respectively. The variance of the flake's stray field ($\Delta$B$\rm ^2_{flk}$) is calculated from the average B$\rm _z$ of the flake. The pixel-wise absolute value of B$\rm _z$ of the flake region, marked by the dashed line in Figs. \ref{flake_1}(a) and (b), was considered to compute the average B$\rm _z$ at a particular temperature. The background magnetic field variance ($\Delta$B$\rm ^2_{bkg}$) is calculated from the region enclosed by the black rectangles in Figs. \ref{flake_1}(a) and (b).
%
%
%
%
 To compute $\Delta$B$\rm ^2_{norm}$, $\Delta$B$\rm ^2_{bkg}$ is subtracted from $\Delta$B$\rm ^2_{flk}$ to remove the background noise followed by appropriate normalization. With an increase in temperature, the vanishing stray-field variance of the flakes indicates the loss of ferromagnetic ordering in 2D FGT flakes as shown in Figs. \ref{flake_1}(c) and (d), complemented by the respective magnetic images. The experimental data can be fitted by a power law of the form $\rm \alpha+\beta*(1-T/T_C)^\gamma$ \cite{zang-PRB-20, chen-ami-23} to extract T$\rm_C$. From the fitted power law in Figs. \ref{flake_1}(c) and (d) the values of T$\rm_C$ are 296.6$\pm$1.5 K and 295$\pm$2.1 K, while  $\gamma$ is 1.78 and 1.48, respectively. This temperature-dependent study expands previous work, and indicates the presence of ferromagnetic ordering and near room-temperature transition temperatures in 2D FGT flakes of varied thicknesses, making this material a suitable candidate for practical spintronic devices.
%
%
\subsection{Thickness Dependence of Transition Temperature and Stray Fields}\label{Bz and Tc}
\begin{figure}
\includegraphics[scale=1]{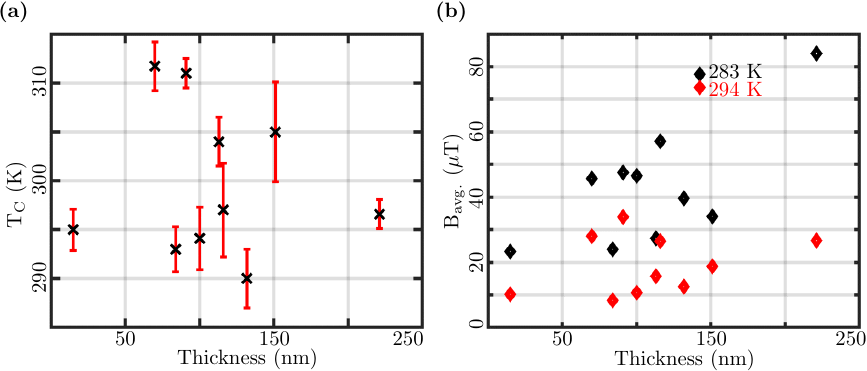}
\caption{(a) T$\rm_C$ verses FGT flake thickness plot. (b) Average magnetic field of an FGT flake as a function of the flake thickness at temperatures 283 K (\ding{117}) and 294 K (\textcolor{red}{\ding{117}}). }
\label{TcVsThickness}
\end{figure}

Based on the capabilities of capturing magnetic images of 2D flakes at varied temperatures with our QDM as described above, we studied the thickness-dependence of the phase transition temperature and the stray fields of the FGT flakes. The T$\rm_C$, for each flake investigated, is extracted following the procedure explained in Sec.-\ref{temp depend}. Following the methods developed for measuring B$\rm _z$ and computing T$\rm_C$, several FGT flakes with varied thicknesses in the range of 15-221 nm were considered. Fig. \ref{TcVsThickness}(a) depicts the observed behavior of T$\rm_C$  for varied FGT flake thickness. This study shows \emph{no dependence} of T$\rm_C$  on the flake thickness in the range of 15-221 nm. These results are consistent with earlier studies, which explored the thickness-dependence of T$\rm_C$  for $\rm Fe_5GeTe_2$ in the range 21-100 nm\cite{chen-IOP-22}. Importantly, our work extends previous results and reports a wider thickness range of the FGT flakes. Also, measured several flakes, analyzing T$\rm_C$ vs. thickness reported as the average thickness of one FGT flake, while earlier studies measured variable thicknesses in a single FGT flake. We note that an anomalous Hall effect-based study of layer number dependent T$\rm_C$  \cite{Deng-NL-22} of $\rm Fe_5GeTe_2$ showed an effect of thickness on T$\rm_C$ in a bilayer and a monolayer, where T$\rm_C$ falls below room temperature.

Moreover, it is worth noting that $\rm Fe_5GeTe_2$ is a ferromagnet, with a near-room temperature T$\rm_C$, down to 15 nm thickness. Fig. \ref{TcVsThickness}(b) shows the B$\rm _z$ as a function of FGT thickness at 283 K and 294 K. Overall, there is an uptrend with the flake thickness. The stray fields (at 283 K \ie below T$\rm_C$)  as a function of thickness plotted in Fig. \ref{TcVsThickness}(b) are similar to the results reported earlier \cite{chen-IOP-22}.

An extensive study of the domain structure as a function of the number of layers of FGT flakes using XPEEM is reported in Ref. \cite{fujita-acs-22}, and T$\rm_C$ for a monolayer was found to be in the range 120-150 K. Comparatively, the domain size of poly-crystalline Mn$_3$Sn films is explored as a function of thickness from a few 10s to 400 nm, indicating an increase with thickness \cite{li-nl-23}.
%
%
%
%
These previous studies showed that the size of the domains critically depends upon the flake thickness, which indeed dictates the overall magnetization of the flake. Earlier results \cite{fujita-acs-22, li-nl-23, chen-ami-23} were obtained on antiferromagnetic materials, at cryogenic temperatures or cobalt-substituted FGT, while the current study is of FGT flakes at ambient temperature. We find that within the examined range of thicknesses (15-200 nm) interfacial and anisotropy effects are not significant, leading to a thickness-independent T$\rm_C$. This is a main result of this work.
%
%
\subsection{Autocorrelation Analysis of Phase Transitions}\label{Autocorrelation_Phasetransition}
We analyzed the evolution of magnetic features in FGT flakes and their $\rm T_C$ as a function of temperature using a normalized 2D correlation function. This analysis provides deeper insights into $\rm T_C$, as well as into the structure of magnetic domains near $\rm T_C$, with an external bias magnetic field in magnetic materials.

The Normalized 2D autocorrelation function can be written as follows.
\begin{equation}\label{autocorrelation function}
\rm \sum_{x,y} \frac{B_{\text{stray}}(x+\delta x, y+\delta y) \cdot B_{\text{stray}}(x, y)}
{\left\| B_{\text{stray}}(x+\delta x, y+\delta y)\| \cdot \|B_{\text{stray}}(x, y) \right\|}.   
\end{equation}
Here, $\rm B_{stray}$ is the stray magnetic field of an FGT flake at pixel (x,y) and $\delta$x and $\delta$y are displacements along the x-axis and y-axis respectively.  The summation quantifies the similarity between the magnetic field at (x,y) to the field at (x+$\delta$x,y+$\delta$y). Hence, this function can be used to find the characteristic scale and behavior of magnetic features and textures of a magnetic image.

\begin{figure}[H]
\includegraphics[scale=0.8]{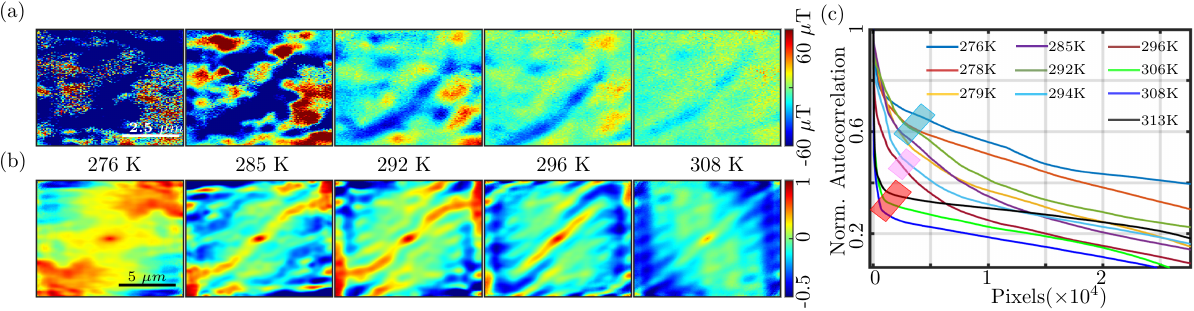}
\caption{(a) Temperature-dependent evolution of magnetic textures in a 100 nm thick FGT flake. (b) Normalized 2D autocorrelation maps, corresponding to the magnetic textures shown in (a), show a change in the spatial correlation as temperature increases. (c) Autocorrelation function, Eq.\ref{autocorrelation function}, of the magnetic textures are plotted against the number of pixels at several temperatures. At low temperatures, the correlation is high and is highlighted by a blue box. As the temperature increases, the correlation starts decreasing highlighted by a pink box. At higher temperatures above T$\rm _C$, the correlation almost vanishes, highlighted by a red box.
} 
\label{Autocorrelation Magnetic}
\end{figure}

Fig. \ref{Autocorrelation Magnetic}(a) shows temperature-dependent magnetic images of a 100 nm thick FGT flake having $\rm T_C$=294.1$\pm$3.2 K. For all magnetic field images of this flake, refer to Supporting Information Section-S6 \cite{si-ACS-25}. At 276 K, the magnetic domains are relatively large, indicating strong ferromagnetic order. As the temperature increases, the size of the domains starts to shrink, reflecting a gradual weakening of the ferromagnetic state due to thermal agitation. Eventually, as the temperature approaches $\rm T_C$, the domains disappear entirely, indicating the transition from a ferromagnetic to a paramagnetic phase. Fig. \ref{Autocorrelation Magnetic}(b) shows the corresponding 2D autocorrelation maps at different temperatures, providing a quantitative perspective on the magnetic domain evolution. At 276 K, the autocorrelation map is nearly uniform with a broad full-width at half maximum (FWHM) around the central peak [($\delta$x,$\delta$y)$=$(0,0)] indicating large and coherent domains. This broad FWHM corresponds to a high degree of correlation across the image, reflecting strong ferromagnetic ordering.

As the temperature increases, the FWHM of the autocorrelation peak decreases, indicating a reduction in the spatial extent of correlated regions. This change is visible in the autocorrelation maps, which show areas of positive correlation (red regions) and negative correlation (blue regions) becoming more localized. The shrinking correlation lengths correspond to the shrinking domain size as thermal energy disrupts the ordered magnetic state. Well above $\rm T_C$, the autocorrelation nearly vanishes, indicating that the material has transitioned into the paramagnetic phase. 
%
%
%
%
In this state, there is minimal correlation between spins across different regions, and the magnetic domains are no longer discernible. These autocorrelation maps provide valuable insights into the evolution of domain sizes with temperature, allowing one to estimate how the magnetic order diminishes as the system approaches and surpasses its $\rm T_C$.

For more quantitative analysis of $\rm T_C$ from the 2D autocorrelation maps, we converted the 2D data to 1D by vectorizing the autocorrelation irrespective of position. The resulting vector was then sorted in descending order and plotted against the pixel number for various temperatures, as shown in Fig. \ref{Autocorrelation Magnetic}(c). From this plot, we observe that at low temperatures, the autocorrelation values are generally high and closely grouped, as indicated by the blue box. This suggests that at lower temperatures, magnetic domains are highly correlated. With increasing temperatures, near $\rm T_C$ the correlation begins to decrease and is reflected in the plot marked by a pink box, and after $\rm T_C$ the correlation value is lowest marked by a red box. 
%
%
\subsection{The Crystallographic Stripe Feature}
A stripe feature in Fig. \ref{flake_1}(a), which becomes clearer at elevated temperatures, is readily observable. A closer inspection of Figs. \ref{flake0}(a) and (b) also reveals similar stripe features in the FGT flakes. The FGT flakes were analyzed using scanning NV center magnetometry (SNVM) and AFM to gain further insights into the stripe feature and their sub-domain magnetic structure at room temperature.

For SNVM, the resolution exceeds the diffraction limit and is controlled by the stand-off distance of the AFM height feedback, typically less than 100 nm. Being closer to the sample surface allows us to resolve finer details of the FGT stray field with a greater signal magnitude while simultaneously carrying out a correlated AFM scan of the sample topography. 
%
%
\begin{figure}
\includegraphics[scale=0.85]{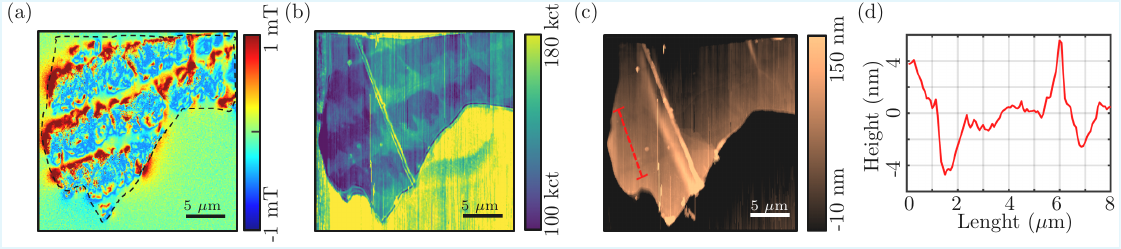}
\caption{(a) Magnetic image obtained using an SNVM showing the stripe features and domain structure. (b) The luminosity plot depicting the photon count rate captured by SNVM tip. (c) The simultaneous topography scan captured by the SNVM tip reveals faint signatures of the stripes. (d) A line-cut profile shown by the red dashed line in (c) indicates the stripes with 3-5 nm depression.}
\label{SNVM}
\end{figure}
%
%

SNVM measurements were carried out on a commercial QZabre QSM system with NV center magnetometry tips obtained from the same supplier. All measurements were performed in an enclosure with temperature regulated at 25$\pm 0.05^{\circ}$ C by feedback. The tip was kept in contact with the sample surface through conventional (lateral) amplitude-modulated AFM feedback on the amplitude of the capacitance of a tuning fork that the diamond probe is attached to. A microwave antenna in the vicinity of the tip along with optical access from the above allowed for an ODMR experiment to be performed at each pixel of a magnetometry scan, thereby obtaining the stray field along the axis of the NV center. A vector magnet module situated directly beneath the sample stage allowed for the application of an arbitrary bias field, typically along the axis of the NV center to pre-split the ODMR peaks. The position of a particular ODMR peak was tracked along the scan. In the presence of strong field gradients, the peak may be temporarily lost, leading to (partial) fit error lines along the course of the measurement. These error lines were excluded from the measurement over the course of routine data analysis.

Results of SNVM measurements on an FGT flake are depicted in Fig.\ref{SNVM}. On a closer inspection of Figs. \ref{flake_1}(a),(b) and Fig. \ref{SNVM}(a) one may observe a quantitative difference in the extracted stray magnetic fields, under similar bias fields, measured by wide-field and scanning NV spectrometers. In order to explore this, further experiments were conducted and results are presented in Supporting Information Section-S7 \cite{si-ACS-25}, indicating consistent measurements. Further, the stripe feature in the stray field shown in Fig. \ref{SNVM}(a) directly coincides with a faint topographical contrast shown in Fig. \ref{SNVM}(c), on the order of 5 nm. The same stripe feature can also be observed in the luminosity plot in Fig. \ref{SNVM}(b). The luminosity plot was obtained from the photon count rate captured by the NV center scanning tip and reveals further details closely related to the magnetic image shown in Fig. \ref{SNVM}(a). The line cut shown in Fig. \ref{SNVM}(d) shows that the stripe region has a depression of the order of 5 nm.

\begin{figure}[H]
\includegraphics[scale=0.95]{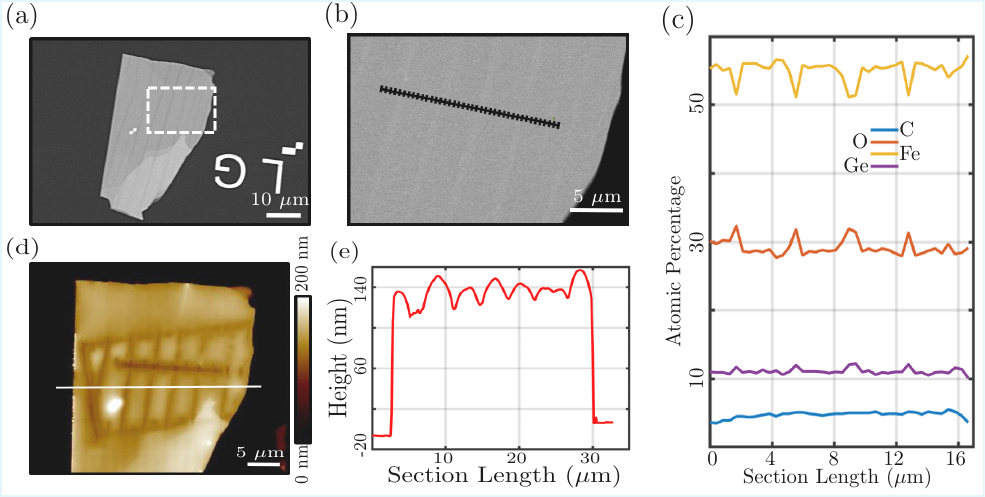}
\caption{(a) A SEM image, at 5000 magnification and 10 keV electron beam energy, of an FGT flake having several stripes. (b) A zoom SEM image of the region enclosed by a white dashed rectangle in Fig (a) at 8000 magnification and 5 keV electron beam energy. The black ladder represents the points considered for the EDXS analysis. (c) EDXS plots, for the region shown by the black ladder in (b), showing the relative atomic percentage of iron, germanium, oxygen, and carbon. (d) The AFM topography of the FGT flake. Regions with additional depressions are the results of the exposure to electron beam during SEM imaging. One may observe even sever damage in the region of EDXS analysis. (e) The line profile of the FGT flake is marked by a white line in (d).}
\label{Oxidation_EDXS}
\end{figure}

For further analysis of the stripe features, energy-dispersive X-ray spectroscopy (EDXS) and AFM measurements were performed. A scanning electron microscope (SEM) image of an FGT flake at 10 keV electron beam energy and 5000 magnification is shown in Fig. \ref{Oxidation_EDXS}(a). One may observe several stripes. 5 keV EDXS measurements were performed on 40 points along a 16.64 $\mu$m line as shown in  Fig. \ref{Oxidation_EDXS}(b). Fig. \ref{Oxidation_EDXS}(c) shows the variations in the atomic percentage for iron, germanium, oxygen, and carbon at the measurement points. Electron beam energy is low for measuring tellurium hence not shown in Fig. \ref{Oxidation_EDXS}(c). Also, a low energy was chosen for EDXS measurements to examine the near-surface region of the FGT flake. In a higher energy electron beam, we observed tellurium but variations in the atomic percentage for other elements were not as prominent as shown in Fig. \ref{Oxidation_EDXS}(c).
%
%
%
%
 One may observe the correlation between the change in concentration of oxygen, iron, and germanium and the respective points on the stripes. These results suggest that there are modulations of iron and germanium concentration in the stripe region. These modulations might be attributed to the respective elemental concentration variations during the synthesis of the FGT. Such modulations caused the surface to oxidize differently compared to the rest of the FGT flake which is further confirmed by the oxygen content on these stripes. 

SEM and EDXS measurements were followed by AFM analysis and the obtained topography is shown in Fig. \ref{Oxidation_EDXS}(d). The e-beam exposed region during EDXS measurement is visible in the AFM image, showing a distinct contrast between the striped regions and the rest of the flake, indicating different interactions of the e-beam with these regions. This could be attributed to the modulation of oxygen, iron, and germanium content in the stripe regions as depicted by the EDXS analysis in Fig. \ref{Oxidation_EDXS}(c). The line profile in Fig. \ref{Oxidation_EDXS}(e), along the white line in Fig. \ref{Oxidation_EDXS}(d), highlights the variation in height across the stripes. Here the depressions are of the order of 10 nm, which seems to be enhanced by the electron beam exposure as compared to the AFM measurement of SNVM Fig. \ref{SNVM}(d). 

It is worth noting that several flakes were examined for the stripe features. The stripe features appears in the magnetic images which were obtained from the surface of the FGT flakes facing the diamond substrate. FGT flakes being encapsulated with gold inside a glove box were not exposed to oxygen (see Supporting Information Section-S5 \cite{si-ACS-25} detailing the effect of oxidation on magnetic image), yet they still exhibit the stripe features in both magnetic and optical images. Hence along with the EDXS and SEM analysis, we conclude that modulation in iron and germanium content during the FGT synthesis, with possibly oxygen also being present, let these stripes oxidize differently leading to our measured results. This might also explain the different FGT behavior in these stripe regions when exposed to a 5-10 keV electron beam, Fig. \ref{Oxidation_EDXS}(d).   
%
%
\begin{figure}
\includegraphics[scale=0.9]{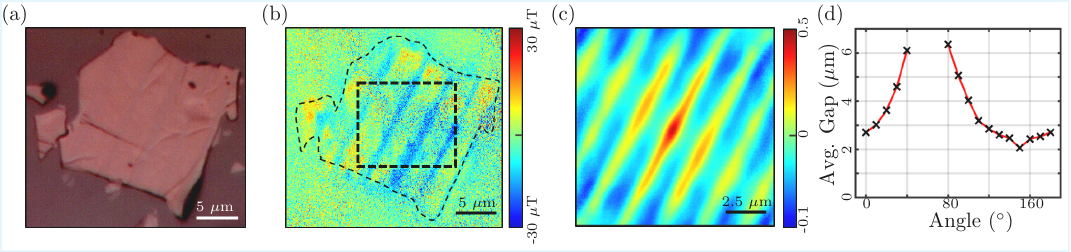}
\caption{(a) Optical image of an FGT flake at 100 magnification showing stripe features. (b) The magnetic stray field-based image of the FGT flake shown in (a) at 307 K. (c) The normalized 2D autocorrelation map of a region marked with a dotted square in (b). (d) Directional 1D autocorrelation analysis of the stripe features from (c), depicting the angular dependence of spatial correlations. The plot illustrates the gap between consecutive stripes increases with the angle and stripes become parallel around 60 degrees. 
} 
\label{Autocorrelation Stripe}
\end{figure}
%
%
Further analysis of these stripe features was carried out by computing the autocorrelation given by Eq. \ref{autocorrelation function}. Fig. \ref{Autocorrelation Stripe}(a) is an optical image of a $\sim$130 nm thick FGT flake. The image shows faint stripe features at 100 magnification. These features are also visible in the magnetic image, Fig. \ref{Autocorrelation Stripe}(b), taken at 307 K of the FGT flake. Fig. \ref{Autocorrelation Stripe}(c) is a normalized 2D autocorrelation map of the region marked with a dashed square in Fig. \ref{Autocorrelation Stripe}(b). The same stripe features are also reflected in the autocorrelation map. This map provides a quantitative analysis of the spatial correlation between the stripes, highlighting the periodicity and micron-order size of these magnetic features in the FGT flake. To further analyze these structures, we extracted angle-dependent 1D autocorrelation data from the 2D autocorrelation map in Fig. \ref{Autocorrelation Stripe}(c). The stripes are fairly consistent in spacing, so the average gap between the stripes is plotted as a function of the angle in Fig. \ref{Autocorrelation Stripe}(d). This plot shows that the gap increases with angle and at $\sim$60 degrees, no gap is detected as around this angle the line along which we are extracting the data becomes parallel to these highly correlated line region. Hence, these features are at $\sim$60 degree angle w.r.t. the horizontal axis with an average gap ranging between 2-3 $\mu$m.

This unexpected crystallographic feature was not previously identified, and our measurements correlate these micron-scale structures to optical, magnetic and elemental signatures. We deduce that these growth features lead to surface oxidation modifications, and as a result to magnetic implications as well. Importantly, while we have shown interfacial effects are not significant in terms of the average $T \rm_C$ of the flakes, such effects, when associated with intrinsic crystallographic features, can lead to local changes in the magnetization distribution. This is another main result of this paper.
%
%
\section{Conclusion}\label{conclusion}
Here, we have demonstrated the capabilities of an NV center based quantum magnetic microscope for non-invasive, high-resolution probing of magnetic features in 2D vdW ferromagnets. The NV center magnetometry technique has allowed us to achieve detailed magnetic field imaging at room temperature and under ambient conditions, overcoming  limitations of traditional approaches that commonly require high-vacuum and/or cryogenic environments.

The quantum magnetic imaging of a cleaved 2D FGT flake revealed sub-micrometer scale magnetic domains and confirmed room-temperature ferromagnetism in the sample. Temperature-dependent measurements showed clear evidence of phase transitions in the FGT 2D flakes at room temperature, with T$\rm_C$  ranging between 285 K to 315 K for thicknesses between 15-221 nm. This study identified no clear thickness dependence of the T$\rm_C$ behavior in $\rm Fe_5GeTe_2$ for the studied thickness range \cite{chen-ami-23, chen-IOP-22, Deng-NL-22}. We also measured the T$\rm_C$  of a bulk single crystal for comparison with exfoliated 2D flakes T$\rm_C$, yet observed no significant variation. Importantly, these results indicate no significant interfacial and anisotropic magnetism effects. Further, the B$\rm _z$ of FGT flake as a function of flake thickness is presented at low (283 K) and high (294 K) temperatures. This study confirms the presence of ferromagnetism in FGT down to 15 nm, making it a suitable candidate for fabricating high-quality 2D magnetic devices at room temperature for spintronic applications. 

We observed a magnetic stripe feature repeatedly appearing during the experiments. It was explored using SNVM, SEM, EDXS, and AFM. We observed that the stripe features can be attributed to modulation in iron, oxygen, and germanium concentrations, which potentially influence surface oxidation, thus leading to these stripes structures. Variations in the oxidation levels were further confirmed by EDXS measurements. We note that these stripe features are not a general feature of FGT flakes but likely result from growth variations, as we did not observe these features in all the available samples. Nevertheless, this structure, which was not previously identified, indicates implications of interface effects on measurable magnetization structures, when accompanied by crystallographic variations.

Moreover, we use high-resolution magnetic imaging and 2D correlation analysis to characterize the magnetic domain structures as a function of temperature. We identify the magnetic spatial behavior across T$\rm_C$ through characteristics of the central lobe of the autocorrelation map. Finally, we identify stripe features in these flakes on the micron scale, which manifest through the magnetic signatures and their 2D correlation features. We thus demonstrate the usefulness of such autocorrelation analysis schemes for characterizing spatial magnetization structures.

As future prospects, using an NV center based quantum magnetic microscope, one can study current-induced domain-wall motion (CIDM), spin-orbit torque, ferromagnetic resonance, spin-transfer torque, magnetic noises, and excitations. Such quantum sensing schemes are specifically well-suited for studying 2D vdW materials \cite{Robertson-IOP-23, johansen-PRL-19, Casola-NRM-18}, to further understand their underlying physics and to evaluate their potential use in spintronic devices and applications.
%
%
\begin{acknowledgement}
The authors thank Prof. Israel Felner (The Racah Institute of Physics, The Hebrew University of Jerusalem, Jerusalem, Israel) for the FGT bulk magnetization measurements and useful discussion. The teams in Mainz and Jerusalem acknowledge funding from the Carl Zeiss Stiftung $($ HYMMS project no.P2022-03-044$)$. A.S. acknowledges the financial support from the Emily Erskine Endowment Fund postdoctoral fellowship. L.C. and A.W. acknowledge funding from the Deutsche Forschungsgemeinschaft (CRC 1552, Project 465145163). M. K. acknowledges support by the Deutsche Forschungsgemeinschaft (DFG, German Research Foundation) projects 403502522 (SPP 2137 Skyrmionics), 49741853, and 268565370 (SFB TRR173 projects A01, B02 and A12), the Horizon 2020 Framework Program of the European Commission under FET-Open grant agreement no. 863155 (s-Nebula) and ERC-2019-SyG no. 856538 (3D MAGiC) and the Horizon Europe project no. 101070290 (NIMFEIA). H.S. is supported by a DFG grant through the SPP Priority Programme (project no. 443404566). N.B. acknowledges support by the European Commission’s Horizon Europe Framework Programme under the Research and Innovation Action GA No. 101070546-MUQUABIS and ERC CoG Project QMAG (no. 101087113). N.B. also acknowledges financial support by the Carl Zeiss Stiftung (HYMMS wildcard), the Ministry of Science and Technology, Israel, the innovation authority (project no. 70033), and the ISF (Grants No. 1380/21 and No. 3597/21). Crystal growth and characterization (AFM) were supported by the U. S. Department of Energy, Office of Science, Basic Energy Sciences, Materials Sciences and Engineering Division. All the authors acknowledge the Harvey M. Krueger Family Centre for Nanoscience and Nanotechnology at the Hebrew University of Jerusalem, Israel for AFM, SEM, and EDXS measurements. 
\end{acknowledgement}
%
%
\begin{suppinfo}
\section*{S1. Sample Preparation}\label{flake_transfer}
Fe$_5$GeTe$_2$ (FGT) bulk crystal was kept inside the Argon-filled glove box ($<$5 ppm of H$_2$O and $<$5 ppm of O$_2$) to prevent oxidation. The thin flakes were mechanically exfoliated from the bulk crystal using silicone-free adhesive plastic tape (Utron Systems Inc., P/N: 1007R-6.0) inside the glove box Fig.\ref{SI_figS1}(a). The sample was repeatedly exfoliated on the tape to get thinner flakes by adhering it to itself Fig.\ref{SI_figS1}(b). These thin flakes were transferred onto the diamond by stamping the tape on the diamond surface Fig.\ref{SI_figS1}(c), with proximity to NV centers. An optical image of transferred FGT flakes onto the diamond surface is shown in Fig.\ref{SI_figS1}{d}.

%
%
\begin{figure}
\renewcommand{\thefigure}{S1}
\begin{flushleft}
    \includegraphics[angle=0,scale=1.06]{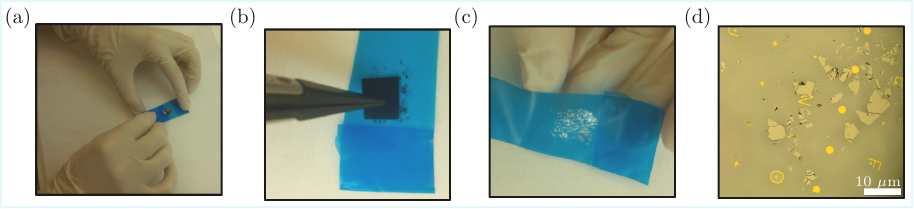}
\end{flushleft}
\caption{(a) Exfoliation of Fe$_5$GeTe$_2$ flakes from the bulk crystal. (b) Thinning the thick flakes transferred from the bulk crystal onto the tape by repeatedly adhering the tape to itself. (c) Stamping the tape for transferring the flakes onto the diamond surface. (d) Optical image of the transferred flakes on the diamond surface having gold markers. Gold markers assist locating the FGT flakes in the wide-field setup.}
\label{SI_figS1}
\end{figure}
%
%
Gold markers were fabricated on the diamond substrate Fig.\ref{SI_figS1}(d) to locate the flakes in the wide-field setup. 10 nm chromium (Cr) was evaporated on the diamond surface as the adhesive layer followed by a 100 nm gold (Au) layer. The markers were printed on these deposited layers by photolithography using a laser writer (LW405B, Microtech, Palermo Italy). This was followed by etching away the unwanted Au and Cr. The diamond was glued to a silicon chip, Fig.\ref{SI_figS1}(b), for easy handling of the diamond during the flake transfer process.

%
%
\section*{S2. Flake Thickness Measurement}
The thickness of the FGT flakes was characterized using Bruker’s dimension XR atomic force microscope (AFM) in tapping mode. The analysis was carried out using Nanoscope analysis software provided by Bruker. Fig.\ref{SI_figS2}(a) depicts an AFM topography, of a FGT flake transferred on the diamond surface. The average thickness is extracted using the step function in Nanoscope analysis software. The step function draws an averaging box with a reference line in the center \ref{SI_figS2}(a) and generates a height profile from the averaged data from both sides of the reference line in the box \ref{SI_figS2}(b). By adjusting the cursor on the height profile average thickness is measured. The average thickness of this particular flake is 137 nm. The thicknesses of all the FGT flakes are extracted using the same procedure.
%
%
\begin{figure}
\renewcommand{\thefigure}{S2}
\includegraphics[angle=0,scale=1.1]{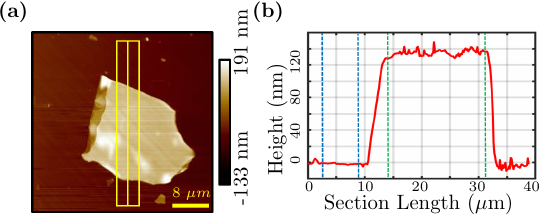}
\caption{(a) Surface topography image of an FGT flake produced by AFM in the tapping mode. (b) The average step-height profile for the selected region shown (yellow) in (a). Here vertical blue and green dashed lines are used to select the length elements (corresponding to an area enclosed by the yellow box in (a)) to estimate the flake height.} 
\label{SI_figS2}
\end{figure}
%
%
\section*{S3. Magnetization Extraction}
Steps followed to extract magnetic field and magnetization images from the pixel-wise ODMR data:
\begin{itemize}
    \item[1.] Pixel-wise, the experimental ODMR data were fitted, and a raw 2D magnetic field image was extracted.
    \item[2.] The average external bias magnetic field (B$\rm ^{ext}_z$) was extracted from the raw image by removing the flake region.
    \item[3.] B$\rm ^{ext}_z$ was subtracted from the raw magnetic image to obtain the magnetic field (B$\rm _{z}$) image of the flake.
    \item[4.] Further, to compute magnetization, the background noises in the magnetic image of the flake were also removed.
    \item[5.] Through reverse propagation the OOP magnetization (M$\rm _z$) of the flake was reconstructed from B$\rm _z$ \cite{tan-IEEE-96,thiel-sci-19}.
    \item[6.] The average magnetization and standard deviation were obtained from the reconstructed M$\rm _{z}$.
\end{itemize}
%
%
\section*{S4. Temperature Control}
A temperature controller (Lightwave LDT-5910B) was used to control the temperature during temperature-dependent measurements. A 14.184 k$\Omega$ thermistor, as the temperature probe, was placed on the diamond surface in the vicinity of flakes. A Peltier-based thermo-electric cooler (Marlow Industries Inc., Model: RC3-6(58954)) was placed on top of the diamond surface for cooling$/$heating along with an aluminum heat sink. A thermal paste (Wakefield Solutions, Type: 120 Silicon) was applied between the diamond surface, thermistor, and cooler for efficient thermal conductivity. During the measurements, the temperature fluctuations were $\sim$$\pm$0.1 K.
%
%
\section*{S5. Thin Flake Chemical Degradation}
In general, iron tends to oxidize when exposed to oxygen. The FGT flakes being composed of iron tend to oxidize when exposed to the air environment. The oxidized FGT layer is no longer ferromagnet and remains on top of the non-oxidized layer, acting as a protective shield for the remaining non-oxidized flake. For thicker layers around 50 nm, the non-oxidized layer retains reasonable thickness and can be measured easily for 3-4 days or longer depending upon the thickness. However, thinner flakes can oxidize completely, making measurement challenging. We also observed the same. A thin flake observed and located in the optical microscope is measured directly after the exfoliation and transferred onto the diamond Fig.\ref{SI_figS5}(a). The same flake was measured after around 12 hours, and we observed that it had decayed, Fig.\ref{SI_figS5}(b). To confirm the flake measured is thin AFM measurement was performed, Fig.\ref{SI_figS5}(c). The AFM analysis reveals its average thickness of around 26 nm, Fig.\ref{SI_figS5}(d).
%
%
\begin{figure}
\renewcommand{\thefigure}{S5}
\begin{flushleft}
    \includegraphics[angle=0,scale=1]{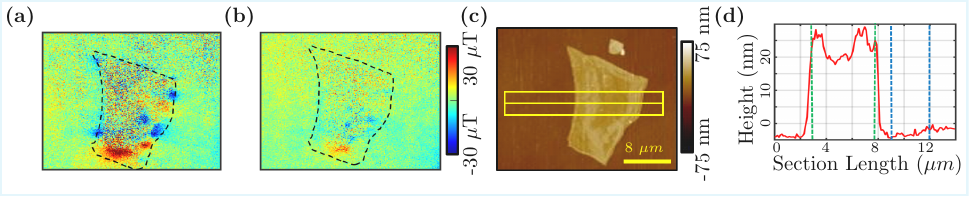}
\end{flushleft}
\caption{Magnetic images of a 26 nm thin FGT flake measured (a) immediately after the transfer onto the diamond surface. (b) 12 hours after the transfer onto the diamond surface. (c) AFM topography image of the FGT flake (d). The average step-height profile was obtained from the yellow rectangle shown in (c).} 
\label{SI_figS5}
\end{figure}
%
%
This oxidation of FGT flakes in an oxygen environment makes it hard to measure thin flakes ($<$20 nm). To prevent oxidation, one can either encapsulate the FGT flake with hexagonal boron nitride flake or evaporate a few nanometers of gold or platinum. Here for measuring thin flakes, the flakes are transferred on the diamond surface in the glove box (MBraun glove box) connected to an evaporator (VST glove box evaporator) and sent directly for evaporating gold $\sim$5 nm to prevent it from oxidation.
%
%
\section*{S6. Temperature Dependent Magnetic Imaging}
\begin{figure}[H]
\renewcommand{\thefigure}{S6}
\includegraphics[angle=0,scale=1.2]{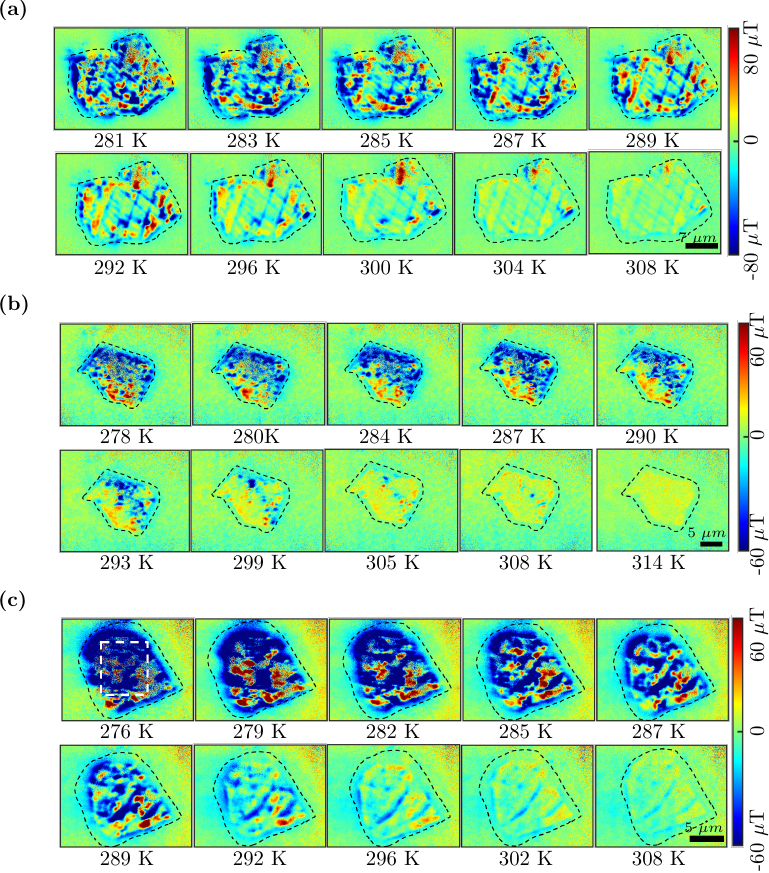}
\caption{(a)-(c) Temperature dependent magnetic images of several FGT flakes. (c) The white dashed rectangle represents the region considered for the autocorrelation map depicted in Fig. 6(b) of the main article.} 
\label{SI_figS6}
\end{figure}
%
%
\section*{S7. Wide-field verses Scanning NV spectrometer magnetic imaging}\label{WFvsSNV}
In order to understand the quantitative difference observed in the computed stray magnetic fields, in Fig.4(a),(b) and Fig.7(a) of the main article, magnetic imaging experiments were repeated on wide-field and scanning NV magnetometers. 
%
%
\begin{figure}[H]
\renewcommand{\thefigure}{S7}
\includegraphics[angle=0,scale=1]{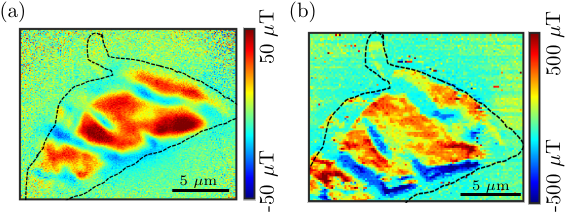}
\caption{Magnetic image obtained from (a) Wide-field NV spectrometer (b) Scanning NV spectrometer.}
\label{SI_figS7}
\end{figure}
%
%
Results are shown in the figure above, \ref{SI_figS7}. These measurements are qualitatively similar, though quantitative differences can be observed in the stray fields. One possible reason could be the difference in the standoff distance of the NV sensors from the FGT flake.
\end{suppinfo}
\bibliography{qs}
\end{document}